# Theory and laboratory tests of the multi-stage phase mask coronagraph

P. Baudoz[a,b], R. Galicher[a,b], J. Baudrand[a,b], A. Boccaletti[a,b]

[a] LESIA, Observatoire de Paris, CNRS, UPMC, Université Paris Diderot; 5 Place Jules Janssen, 92190 Meudon, France
[b] Groupement d'Intérêt Scientifique PHASE (Partenariat Haute résolution Angulaire Sol Espace) between ONERA, Observatoire de Paris, CNRS and Université Paris Diderot.

**ABSTRACT**

A large number of coronagraphs have been proposed to overcome the ratio that exists between the star and its planet. The planet finder of the Extremely Large Telescope, which is called EPICS, will certainly need a more efficient coronagraph than the ones that have been developed so far. We propose to use a combination of chromatic Four Quadrant Phase Mask coronagraph to achromatize the dephasing of the device while maintaining a high rejection performance. After describing this multi-stage FQPM coronagraph, we show preliminary results of a study on its capabilities in the framework of the EPICS instrument, the planet finder of the European Extremely Large Telescope. Eventually, we present laboratory tests of a rough prototype of a multi-stage four-quadrant phase mask. On one hand, we deduce from our laboratory data that a detection at the $10^{-10}$ level is feasible in monochromatic light. On the other hand, we show the detection of a laboratory companion fainter than $10^{-8}$ with a spectral bandwidth larger than 20%.

## 1. INTRODUCTION

The recent indirect discovery of planetary objects orbiting around a large number of nearby stars has motivated the development of instrumental techniques to perform direct detection of these exoplanets. The direct spectroscopic study of the few photons emitted or reflected by the atmosphere of the planet appears as the best solution to fully understand the planet formation and evolution. However, direct detection is quite challenging, since planets are much fainter and angularly close to their host star. Coronagraphy is one of the promising techniques that could efficiently contribute to planet direct detection and characterization by removing or attenuating a large part of the stellar light. Among those techniques, our group has developed an expertise on a phase-mask coronagraph proposed by Rouan et al. 2000 [1], called the four-quadrant phase-mask (FQPM). The focal plane is split into four equal areas, two of which are phase-shifted by π. As a consequence, a destructive interference occurs in the relayed pupil, and the on-axis starlight rejected outside the geometric pupil is filtered with an appropriate diaphragm called the Lyot stop. The capabilities of the FQPM have been studied from a theoretical point of view (Rouan et al. 2000 [1]; Riaud et al. 2001 [2]) and in the lab (Riaud et al. 2003 [3]). Our group at the Observatoire de Meudon is also participating in several projects related to the direct imaging of exoplanets. First of all, in agreement with ESO, we have installed a near infrared FQPM inside the Adaptive Optics camera NACO (Rousset et al. 2003 [4]) on UT4 at the VLT in 2004 (Boccaletti et al. 2004 [5]). In 2007, we have upgraded the coronagraphic capability of NACO by installing two new FQPM. One is working in the K band and has an improved efficiency compared to the previous one thanks to a very small Lyot mask in the center 0.15" in diameter, which mitigates the effect of residual jitter. The other FQPM recently installed on VLT is optimized to be used with the Spectral Differential Imaging Instrument (SDI Lenzen et al. 2004 [6]), which is associated with NACO. Our group is also responsible for the study, manufacture, characterization, and delivery of a set of coronagraphic devices for several instruments: the mid-IR imager of the James Webb Space Telescope (Baudoz et al. 2006 [7]), the SPHERE instrument for the VLT (Boccaletti et al. 2008 [8]).

In the context of EPICS, the planet finder that will take place on the European Extremely Large Telescopes (E-ELT), the detection of large rocky planets called super-earths is arising as a major scientific goal (Kasper et al., 2008 [9]). To reach such an ambitious goal, the study and development of coronagraph more efficient than the one developed so far is mandatory. Measuring the few photons reflected by the atmosphere of faint planets like super-earths will require to reach very high contrasts over broad spectral bandwidth. Theoretically and practically, a phase mask coronagraph is able to reach very high contrasts. However, in its simplest versions, such a coronagraph is chromatic since the π phase shift is performed for a single optimized wavelength. We propose to use a combination of chromatic phase mask coronagraphs in cascade to achromatize the overall dephasing of the coronagraph. For our analysis and our test, we used the FQPM because its performance as a single coronagraph is well studied.

## 2. PRINCIPLE AND THEORETICAL PERFORMANCE

To increase achromaticity of the coronagraph, we propose to use several monochromatic coronagraphs in cascade. This is possible because the light that propagates through a single FQPM can be separated into two parts. The part of the light that interfere destructively which is diffracted outside of the pupil geometry in the pupil plane downstream of the coronagraph. Another part of the light appears not affected by the coronagraph and cannot be blocked out by a Lyot stop. For a monochromatic FQPM working at its optimized wavelength and with a full circular pupil without phase defects, all the light is diffracted outside of the pupil geometry (Abe et al. 2003 [10]). A Lyot stop with the exact size of the pupil can completely block this light. Departing from the optimized wavelength, there will be a fraction of the light that does not interfere. In the extreme case of an observing wavelength equal to half of the optimized wavelength, the dephasing of the coronagraph is 2π and there is no diffracted light outside of the pupil geometry (Fig. 1).

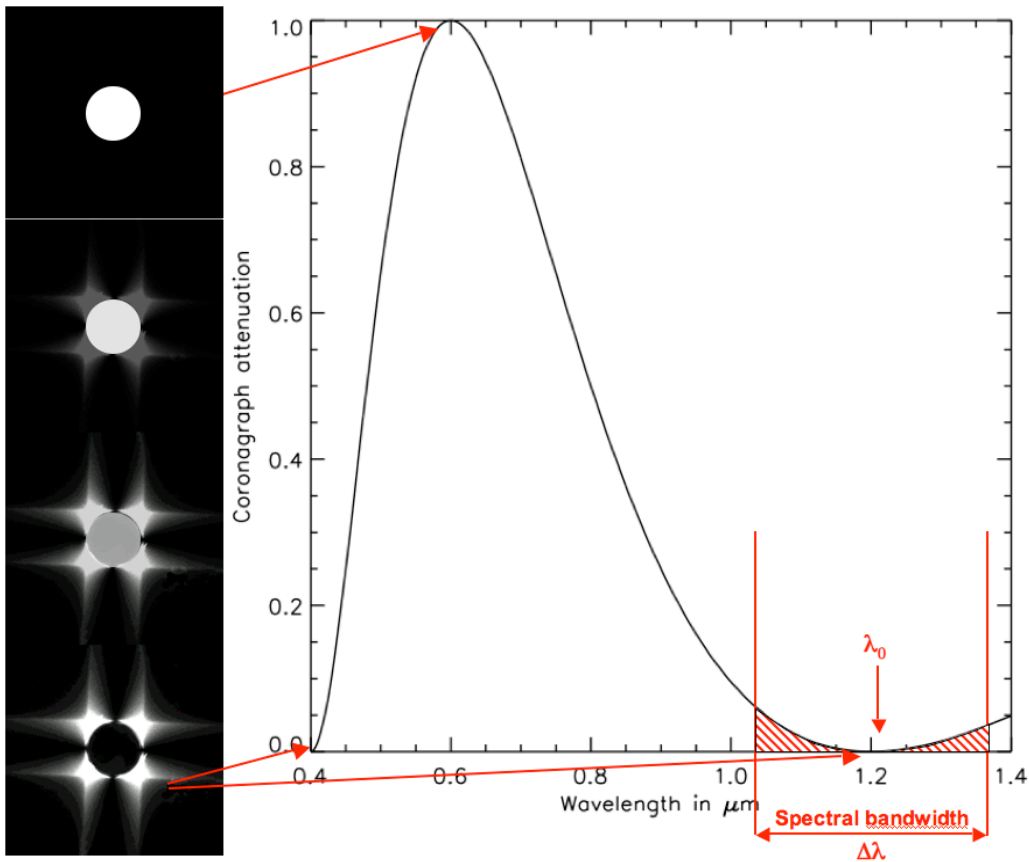

Fig. 1: Theoretical chromatic dependency of a typical monochromatic FQPM. The optimized wavelength of the FQPM is chosen at 1.2 μm by introducing a simulated optical thickness on two quadrants out of four. Assuming no dispersion of the substrate, there should be no coronagraphic effect at 0.6 μm and a second order null at 0.4 μm

## 2.1 Chromaticity

Since the amplitude distribution of light that does not destructively interfere is not modified in the pupil plane downstream of the coronagraph, we can propagate the beam in another FQPM coronagraph that has a different optimized wavelength. This could be done infinitely to increase the achromaticity of the coronagraph. With the number of optical surfaces (stages) increasing, the problems related to alignment and non-perfect optics will also increase and a trade-off analysis has to be carried out depending on the expected performance and the system complexity. Depending on the number of implemented stages, the wavelength of each FQPM must be optimized to efficiently attenuate the entire spectral bandwidth. For a single FQPM and for a perfect wavefront, Riaud et al 2003 [3] demonstrated that the total rejection rate (Ratio of total intensity of the direct image to that of the coronagraphic image) depends only on spectral bandwidth as : $48/\pi^2 \cdot (\lambda_0/\Delta\lambda)^2$. As we add more stages, simulation shows that the rejection rate varies roughly as $[48/\pi^2 \cdot (\lambda_0/\Delta\lambda)^2]^N$ with N the number of stages (Fig. 2). For a typical astronomical bandwidth ($R=\lambda_0/\Delta\lambda=5$), a rejection rate of $10^8$ is expected with a multistage coronagraph using only 4 FQPMs.

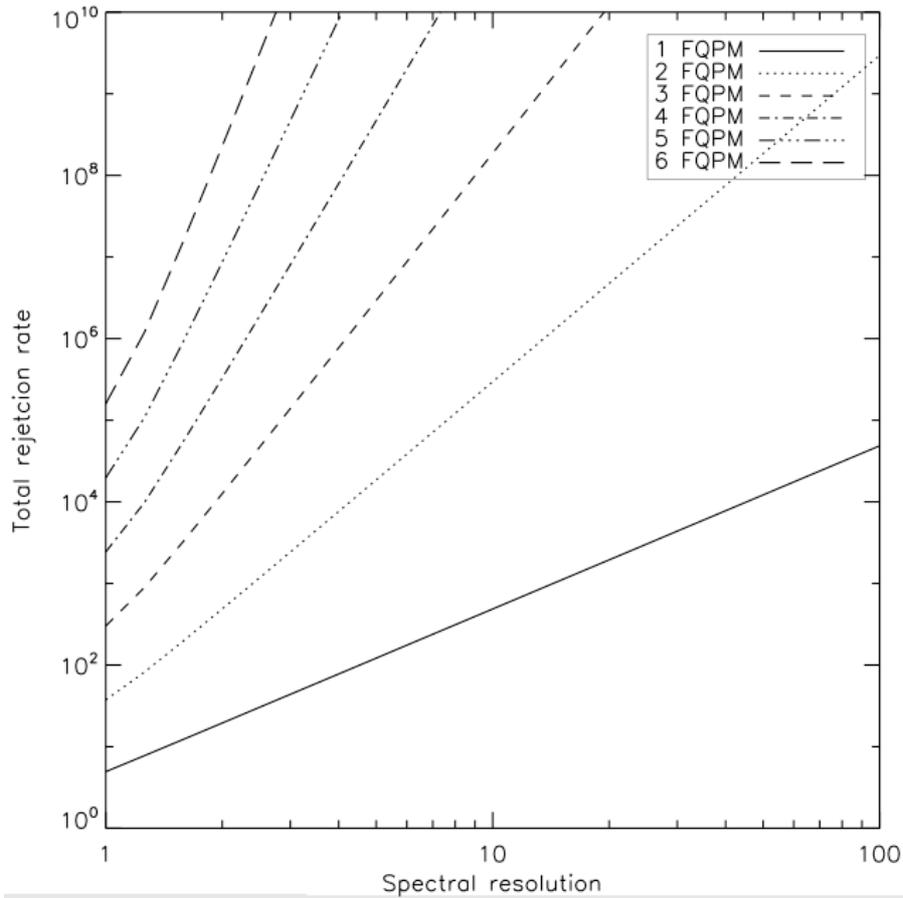

Fig. 2: Total rejection rate (Ratio of total intensity of the direct image to that of the coronagraphic image) as a function of the spectral resolution $R=\lambda_0/\Delta\lambda$. The different curves show results from single FQPM to a multi-stage FQPM using 6 coronagraphs

## 2.2 Stellar Leakage

A single FQPM coronagraph is very efficient to observe at a short angular distance of the star, but at the same time is high sensitive to the low-order aberrations and stellar angular size. The multi-stage FQPM mitigates this problem because, at each stage, a single FQPM has a capacity to decrease the intensity from a source that is not completely outside the inner working angle (IWA), ie the area where a point source is decreased by a factor of at least 50%. Using 4 FQPMs in cascade enables to decrease by a factor 10 the sensitivity of the coronagraph to stellar angular size (Fig. 3). Observing in H band on an E-ELT of 42 m diameter, the total rejection rate of the coronagraph is above $2.10^4$ for a sun-like star located at 20 pc. This level of $2.10^4$ has to be compared to other limitations that arise from the telescope pupil shape or from the uncorrected atmospheric turbulence.

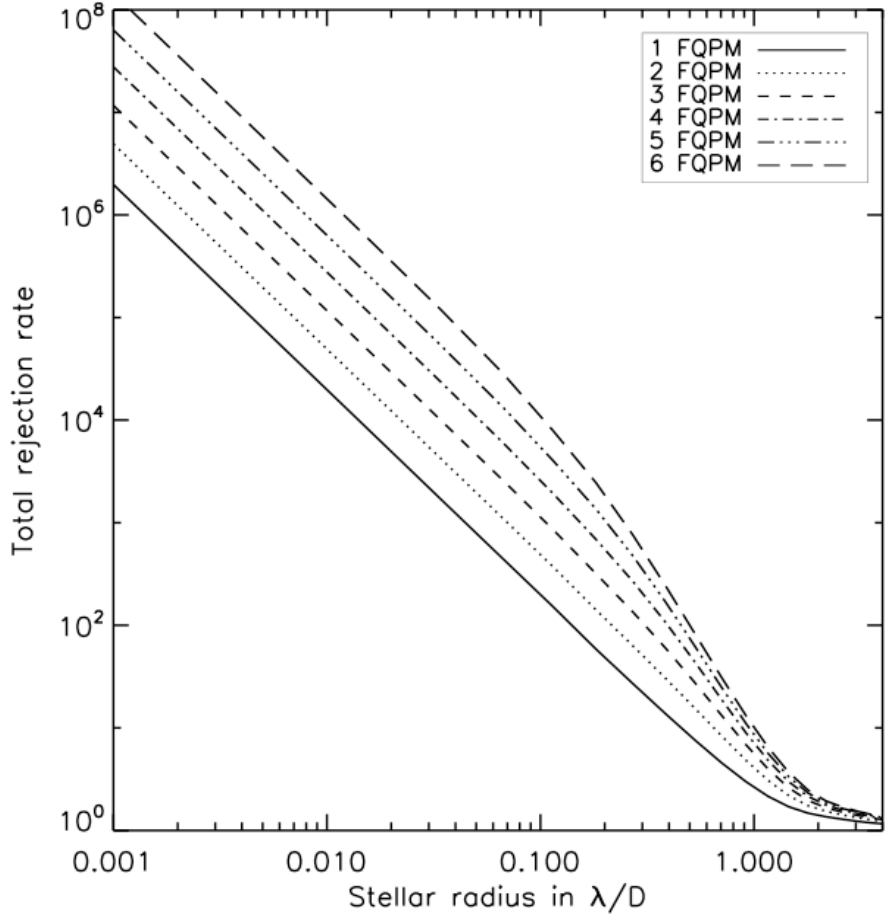

Fig. 3: Total rejection rate as a function of the angular separation of a point source to the multi-stage coronagraph optical axis. The rejection rate reaches 1 (no effect from the coronagraph) for stellar radius larger than 2 $\lambda/D$

**2.3 Central obscuration**

The performance of the FQPM is also degraded when the pupil is not anymore a full circular pupil. All the structures that block the light at the level of the entrance pupil introduce diffraction effects in the Lyot plane. The distribution of light in this plane is concentrated around the edges of these structures (central obscuration, spider, segment gap). The shape of the Lyot stop can then be optimized. To do so, we used the same metric than Boccaletti et al. 2004 [11], which tries to maximize the ratio of the planet throughput to the stellar diffraction residuals. We studied the optimization of each Lyot stop by measuring its impact on our metric. Preliminary results are presented here based on the specific case of the E-ELT study for the EPICS instrument assuming a coronagraph made of four stages of FQPM is used. Two spider structures have been studied (pupil 1 and pupil 5 case, see Fig. 5). We found that the best shape for the first Lyot stop is a shape exactly equal to the entrance. The following stops ($2^{nd}$, $3^{rd}$ and $4^{rth}$ Lyot stops) are optimized to be all the same and their shape is shown in see Fig. 5.

The Fig. 4 shows the variation of the Lyot stop optimization metric for two E-ELT pupil designs (Fig. 5) and for two ranges of angular area used for the estimation of the stellar diffraction residuals. On one hand we compare the planet throughput to the diffraction residual integrated for all radial distances between 1 and 100 λ/D and on the other hand the planet throughput is compared to the diffraction residual integrated only between 5 λ/D and 40 λ/D. The pupil 5 is better optimized for MS-FQPM (factor of 2) for both cases. The integration from 1 to 100 λ/D does not show a clear optimum value while the integration over 5 to 40 λ/D peaks for both design at a transmission of the pupil of 88%.

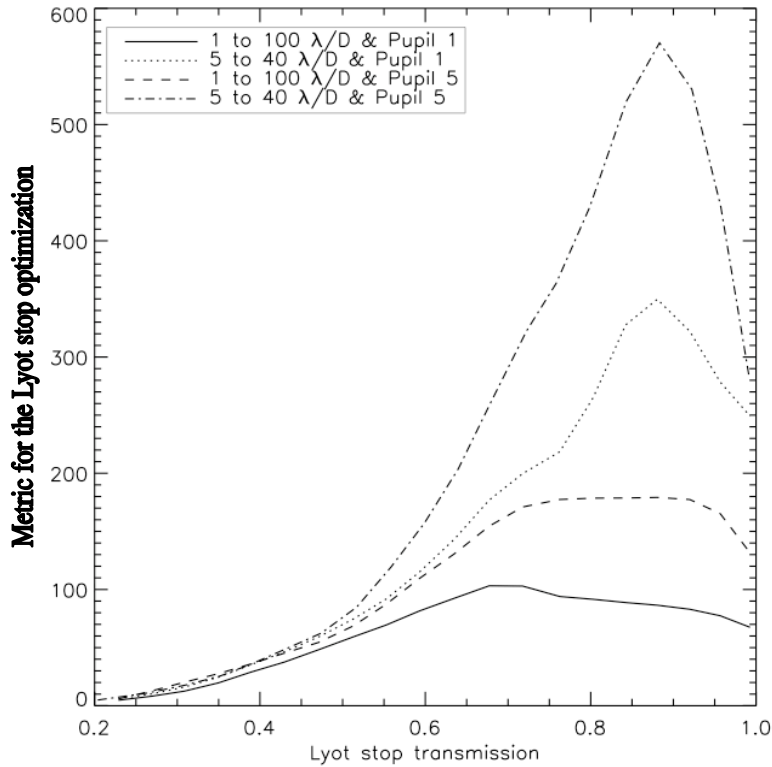

Fig. 4 : Value of the metric (troughput divided by stellar diffraction residual) for two pupil designs and two cases of stellar diffraction residual: 1 to 100 λ/D and 5 to 40 λ/D.

The shape of the optimal Lyot stop of 88% is shown in Fig. 5. In theory, the shape of the stop depends on the orientation of the quadrants of the mask with respect to the orientation of the spider vanes. In the case modeled here the cross lines of the mask are vertical and horizontal for the orientation of the picture.

In practice, the change of the orientation has no impact on the Lyot stop transmission since the residual light that is diffracted in the pupil mostly comes from the central obscuration. As shown in Fig. 4, a change of orientation of the spider decreases the efficiency of the Lyot coronagraph at a given transmission but does not change the best Lyot transmission for a given set of spider. However, the optimized shape of the Lyot stop depends on the spider shape (Fig. 5). The metric we use only measures the stellar diffraction residuals. This is not optimal and it may have to be redone later with more detailed assumptions on the static and dynamic aberrations. This is especially true because the final raw contrast is limited by residual speckle instead of diffraction. Several aspects related to the coronagraph are also neglected in this basic analytic simulation: the dispersion of the substrate on which the FQPM can be fabricated and the fabrication defects of the coronagraph.

Using the optimized Lyot calculated above, we calculated the azimuthally average contrasts given by a four-stage FQPM (Fig. 6). The pupil 5 design gives better average profiles, especially between 10 and 30 λ/D where the E-ELT will search for large exoplanets. This level is of the same order than the limitation arising from the stellar leakage ($2.10^4$) of nearby stars. A more detailed study needs to be performed to evaluate the mixing of these effects to the limitation due to the residual atmospheric turbulence not corrected by adaptive optics.

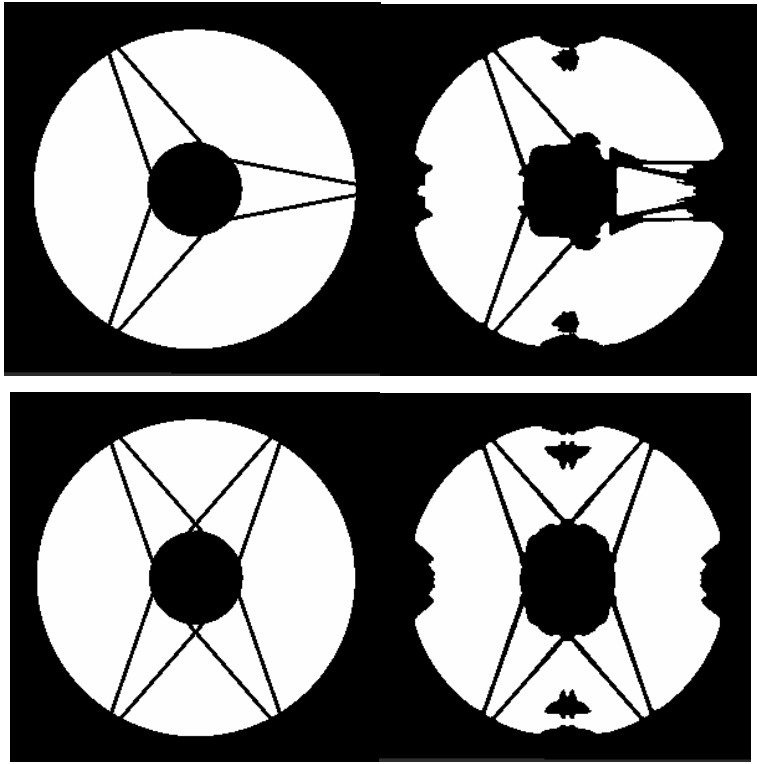

Fig. 5. Left column: Pupil shape and 1st Lyot stop shape (which is equal to the pupil shape, see text) for two possible E-ELT design. Right column: $2^{nd}$, $3^{rd}$ and $4^{rth}$ optimal Lyot stop shape. Upper: pupil design 1. Bottom: pupil design 5.

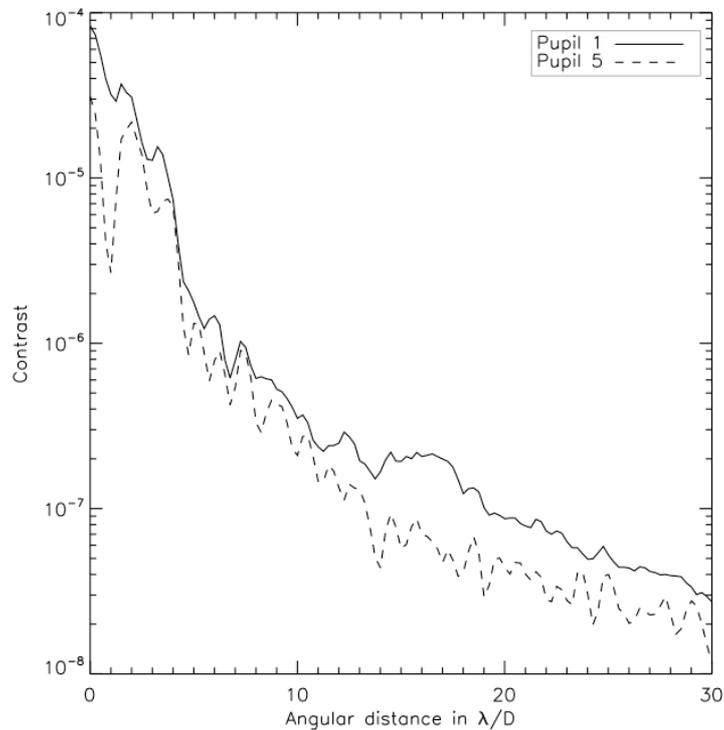

Fig. 6 : Average profile of the MS-FQPM with 4 stages with the optimized Lyot stops for two E-ELT pupil designs. Perfect wavefront, monochromatic simulation and perfect point-like source.

# 3. LABORATORY TESTS

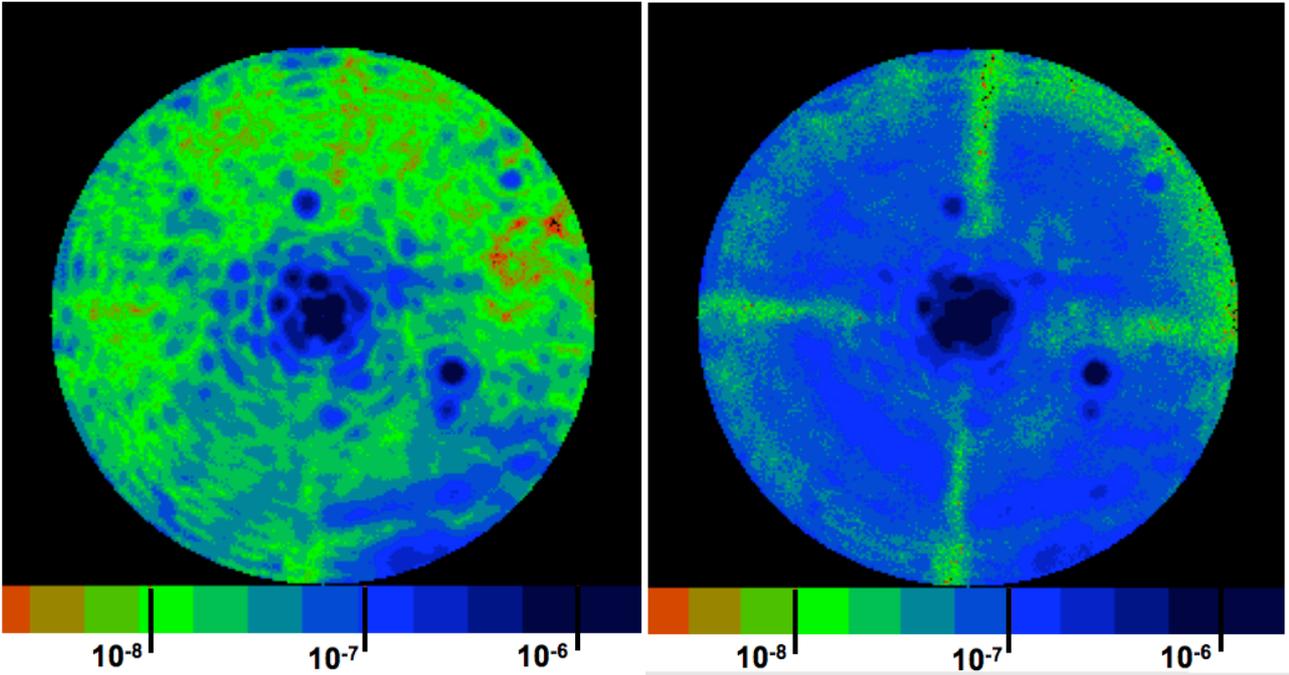

Fig. 7 : Image obtained on the bench of a MS-FQPM made of 3 FQPM centered around 650nm. Left : $\lambda_0/\Delta\lambda$=11. Right : $\lambda_0/\Delta\lambda$ =4.7.

A simple prototype was developed to test the principle of MS-FQPM with 3 stages. We used visible FQPM centered around 635 nm. This was not optimized to achromatize the attenuation. We tested spectral resolutions of R=$\lambda_0/\Delta\lambda$=11 and R=4.7 and monochromatic light (laser diode with less than 1nm bandwidth). The entrance pupil size was 1.78 mm and the Lyot stops for the 3 stages were respectively 77%, 82%, 73% with an overall 21.7 % of the entrance pupil for the final Lyot stop. The contrast obtained at 3 $\lambda/D$ is better than $10^7$ in all cases (Fig. 7). The bright spots visible in the images are ghosts created by AR coatings that were not optimized for this prototype. In the next two sections, we derive the detectability level reachable with the data recorded on out test bench.

## 3.1 Monochromatic detectability

To compare with other such detectability estimation (Trauger &Traub 2007 [12]), we take the same assumption of a space telescope observation where the whole instrument rotate. We added a 15° rotation between each of our images and used a dedicated program that removes the median image to each of the images. We used a data set that last approximately 10 minutes with a total of 400 monochromatic images that can decrease the RON to 1 e- per pixel. The detectability is derived from the final processed image by calculating for each angular distance the residual RMS value as described in Boccaletti et al. 2004 [11]. From our preliminary results, a $10^{-10}$ level is reached at a distance of 10 $\lambda/D$ on our test bench in monochromatic images. On the contrary to Trauger & Traub 2007 [12], our experiment is not stable over a long time because we are not in the vacuum and there is no deformable mirror implemented on our bench. To verify our data processing, we added numerically a faint planet to our data. The planet is a copy of the off-axis image of the source but multiplied by a factor $5.10^{-10}$ and shifted by 7.6 $\lambda/D$ (Fig. 8).

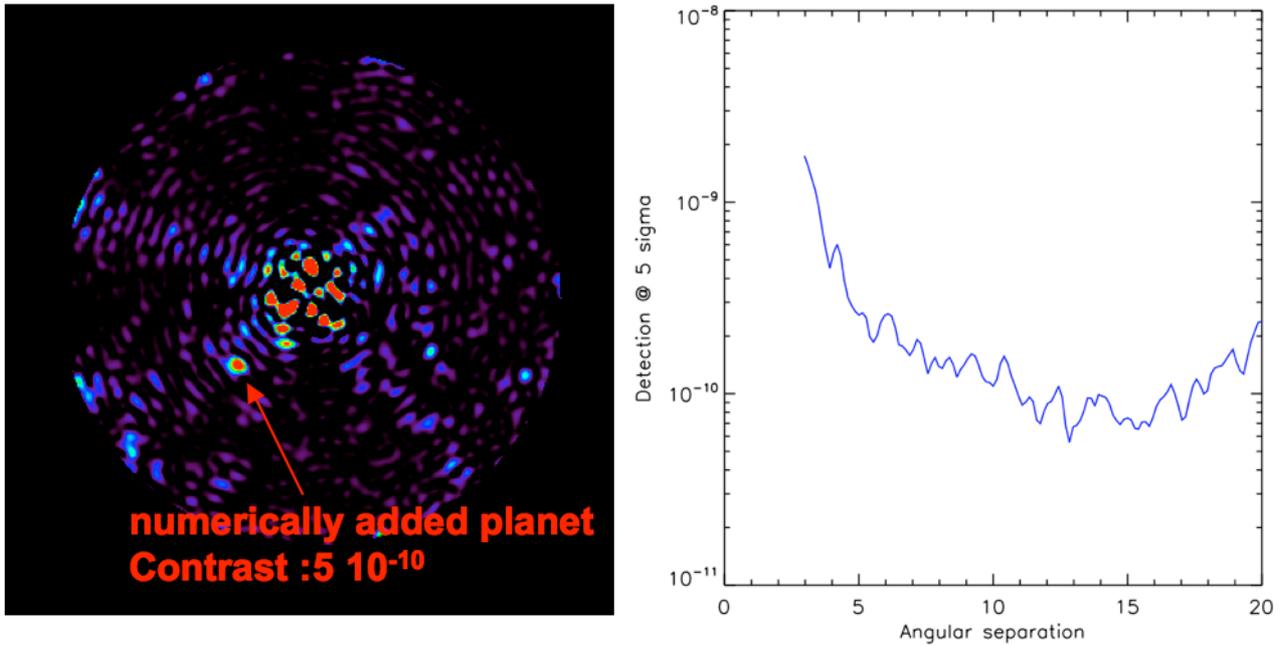

Fig. 8 : Left: Processed image showing the detection of a numerically added planet assuming rotation of the planet compared to the speckle field. Right: 5 sigma detection as a function of angular separation in λ/D calculated with the same hypotheses. Monochromatic light for both data.

## 3.2 Polychromatic detectability

We also tried to estimate the detectability with larger spectral bandwidth using a short data set of less than a minute. We modified our optical setup to introduce a laboratory planet using a beamsplitter close to the entrance pupil plane. The star source was created using a continuum source created by 10 picosecond pulses in a nonlinear supercontinuum generator. The spectral bandwidth of this fibered source is spanning from 450nm to beyond 1750nm and we chose a shorter bandwidth by the mean of an optical filter placed in front of the CCD detector. The planet source was a simple halogen bulb light injected in a single-mode fiber. The detection capability is measured between two sets of measurements recorded with and without the planet source powered on. A simple subtraction of both sums of 20 images (Fig. 9, left) clearly shows the detection of this laboratory planet which has a flux ratio of $6.7\ 10^{-9}$ with the main source for a spectral bandwidth of 130 nm ($\lambda/\Delta\lambda=4.8$). The distance of this laboratory planet to the main source is only 4.5 λ/D. We also measure the detectability of a faint source achieved after a simple subtraction image using two data sets without laboratory planet (Fig. 9, right). These results are very encouraging for tests made on typical laboratory environment without deformable mirror.

The beam aperture for these tests was not optimized for a real instrument (F/D=300). In the framework of the EPICS instrument for the E-ELT, we have started the development of a prototype optimized in the visible wavelength and for faster beam (F/D=40). The main technical difficulties come from the fabrication of sharp transition for the phase step. This is a common problem for all phase coronagraphs. This is mitigated by the multiple stage effect but is still an issue for a fast beam especially at short wavelength. Another issue is the precise positioning and the stability of each of the coronagraphs on the same optical axis.

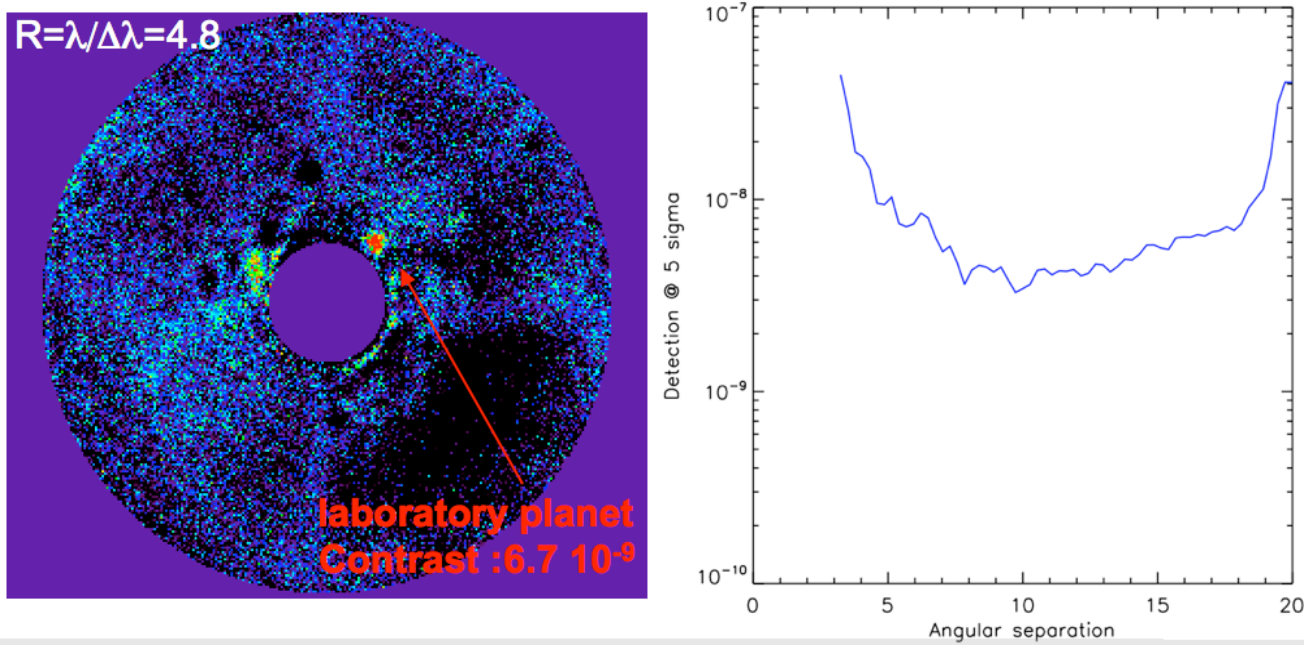

Fig. 9 : Left: Processed image showing the detection of a real laboratory companion source assuming a reference image without companion is subtracted afterward. Right: 5 sigma detection as a function of angular separation in l/D calculated with the same hypotheses. Polychromatic results with a spectral bandwidth of 130 nm.

## 4.  CONCLUSION

After a brief description of the principle of the multiple-stage FQPM coronagraph, we have shown preliminary study on its capabilities in the framework of the EPICS instrument, the planet finder of the European Extremely Large Telescope. From laboratory tests of these multi-stage four quadrant phase mask, we deduce that a detection at the $10^{-10}$ level is feasible in monochromatic light at 10 λ/D. We have also performed the actual detection of a laboratory companion fainter than $10^{-8}$ located at 4.5 λ/D and with a spectral bandwidth larger than 20% . Following these very encouraging results, we are developing a compact prototype with a F/D=40 in the framework of the EPICS instrument.

## REFERENCES

[1] Rouan, D., Riaud, P., Boccaletti, A., Clénet, Y., Labeyrie, A. 2000, PASP, 112, 1479
[2] Riaud, P., Boccaletti, A., Rouan, D., Lemarquis, F., Labeyrie, A. 2001, PASP, 113, 114
[3] Riaud, P., Boccaletti, A., Baudrand, J., Rouan, D. 2003, PASP, 115, 712
[4] Rousset, G., et al. 2003, Proc SPIE 4839, 140
[5] Boccaletti, A., Riaud, P., Baudoz, P., Baudrand, J., Rouan, D., Gratadour, D., Lacombe, F., Lagrange, A.-M., 2004, PASP, 116, 1061
[6] Lenzen, R., Close, L., Brandner, W., Hartung, M., & Biller, B., 2004, Proc. SPIE 5492, 970
[7] Baudoz, P. Boccaletti, A., Riaud, P., Cavarroc, C., Baudrand, J., Reess, J.-M., Rouan, D. 2006, PASP, 118, 765
[8] Boccaletti, A., et al., 2008, This SPIE volume.
[9] Kasper, M., et al., 2008, this volume.
[10] Abe, L., Domiciano de Souza, A., Vakili, F., Gay, J. 2003, A&A, 400, 385
[11] Boccaletti, A. 2004, Astronomy with High Contrast Imaging II, EAS Publications Series, 12, 165
[12] Trauger, J., Traub, W. 2007, Nature, 7137, 771